\documentclass[onecollarge,natbib]{svjour2}
\bibpunct{[}{]}{;}{n}{}{,} % to get "[numbered]" references from natbib
\smartqed  % flush right qed marks, e.g. at end of proof
\usepackage{graphicx}
\usepackage{amsfonts,amsmath,amssymb,bm}

\newcommand{\be}{\begin{equation}}
\newcommand{\bea}{\begin{eqnarray}}
\newcommand{\ee}{\end{equation}}
\newcommand{\eea}{\end{eqnarray}}
\newcommand{\sla}{\slash \hspace{-0.22cm}}

\def\s#1{{\scriptscriptstyle #1}}
\def\noeq#1{(\ref{#1})}
\def\1eq#1{Eq.~(\ref{#1})}

\def\2eqs#1#2{Eqs.~(\ref{#1}) and~(\ref{#2})}
\def\3eqs#1#2#3{Eqs.~(\ref{#1}),~(\ref{#2}) and~(\ref{#3})}
\def\Q#1{Q_{#1}}
\def\oQ#1{\overline{Q}_{#1}}

\def\fig#1{Fig.~\ref{#1}}

\journalname{Few-Body Systems}
\begin{document}

\title{Hadron phenomenology from first-principle QCD studies}


\author{Joannis Papavassiliou}


\institute{Department of Theoretical Physics and IFIC, 
University of Valencia and CSIC,
E-46100, Valencia, Spain \\
              \email{Joannis.Papavassiliou@uv.es}  
              }
\date{Received: date / Accepted: date}

\maketitle

\begin{abstract}

The form of the kernel that controls the dynamics of the Bethe-Salpeter equations is 
essential for obtaining quantitatively accurate predictions 
for the observable properties of hadrons. In the present work  
we briefly review the basic physical concepts and field-theoretic techniques employed in a first-principle
derivation of a universal (process-independent) component of this kernel. This ``top-down'' approach 
combines nonperturbative ingredients obtained from lattice simulations and Dyson-Schwinger equations,   
and furnishes a renormalization-group invariant quark-gluon interaction strength, which is 
in excellent agreement with the corresponding quantity obtained from a systematic  ``bottom-up'' treatment,  
where bound-state data are fitted within a well-defined truncation scheme.

\end{abstract}

\section{Introduction}
\label{intro}

\begin{figure}[!t]
\includegraphics[scale=.6]{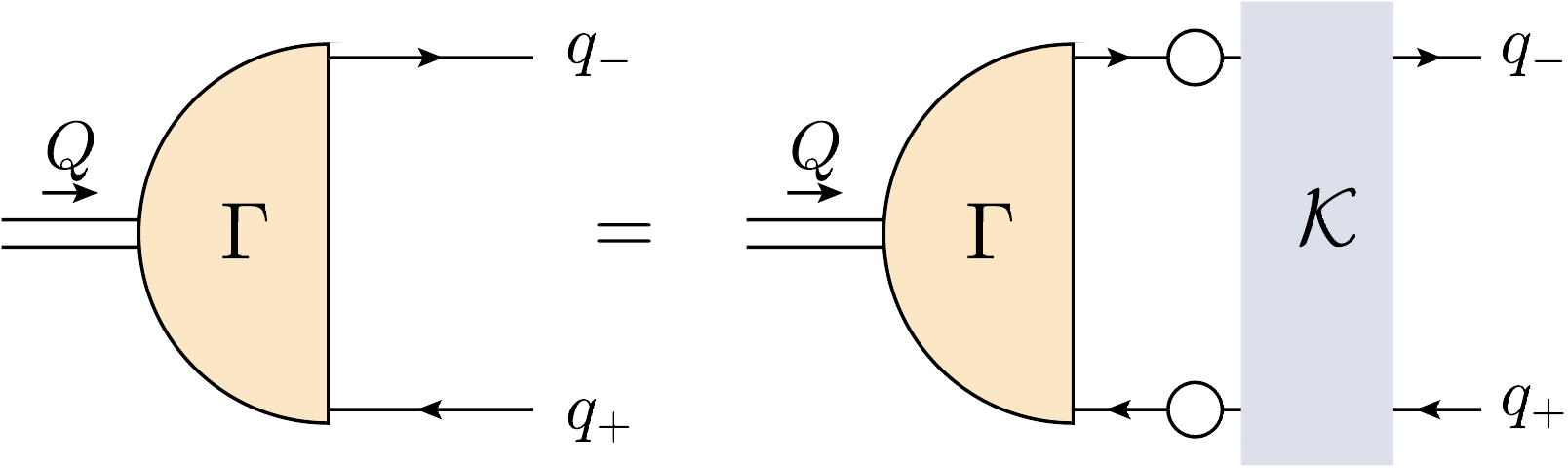}
\caption{\label{fig:BSkernel}The meson BSE and the kernel ${\cal K}$.}
\end{figure}

The spectrum and various physical properties 
of the mesons are  traditionally obtained in the continuum 
by means of special integral equations, known as Bethe-Salpeter equations 
(BSEs)~\cite{Jain:1993qh,Munczek:1994zz,Bender:1996bb,Maris:1997hd,Bender:2002as,Chang:2009zb,Fischer:2009jm,Eichmann:2009zx,Eichmann:2008ef}, 
whose general form is captured by the diagram in Fig.(\ref{fig:BSkernel}).
In this particular eigenvalue equation, ${\rm \Gamma}$ denotes the so-called Bethe-Salpeter amplitude, 
and  ${\cal K}$ the fully-amputated  quark-antiquark scattering kernel. 
The details of the solutions obtained from BSEs depend crucially on the precise form of ${\cal K}$,
and the nonperturbative information included in it. In fact, it is well-known that any 
self-consistent analysis based on BSEs must be intimately connected with 
nonperturbative phenomena such as chiral symmetry breaking, and quark and gluon 
mass generation, which are described by the dynamical equations 
obeyed by the Green's functions of the theory, namely the 
Dyson-Schwinger equations (DSEs)~\cite{Roberts:1994dr,Maris:2003vk}. 
In this presentation we briefly review recent work that aims at a first-principle 
derivation of a special component of the BSE kernel~\cite{Binosi:2014aea}, 
and compare the results obtained with phenomenologically successful ``bottom-up'' versions of the same quantity.

\section{Definitions and basic ingredients}

In the Landau gauge
the gluon propagator is given by  
\be
i\Delta_{\mu\nu}(p)=- i\left[ g_{\mu\nu}- p_\mu p_\nu/p^2\right]\Delta(p^2)\,,
\label{Delta}
\ee
while the ghost propagator, $D(p^2)$, and its dressing function, $F(p^2)$,
are related by $D(p^2)= F(p^2)/{p^2}$.

In addition, consider a special  
two-point function, denoted by $\Lambda_{\mu\nu}(p)$, defined as
\bea
 \Lambda_{\mu \nu}(p) &=&g_{\mu\nu} -i g^2C_A
\int_k H^{(0)}_{\mu\rho}
D(k+p)\Delta^{\rho\sigma}(k)\, H_{\sigma\nu}(-k-p,k,p),
\nonumber \\
&=& g_{\mu\nu} [1+ G(p^2)] + \frac{p_{\mu}p_{\nu}}{p^2} L(p^2);
\label{LDec}
\eea
where $C_{A}$ is the Casimir eigenvalue of the adjoint representation, 
and \mbox{$\int_{k}\equiv\mu^{2\varepsilon}(2\pi)^{-d}\int\!d^d k$}, 
with $d=4-\epsilon$ the dimension of space-time. 
The quantity $H_{\mu\nu}$ corresponds to the well-known ghost-gluon kernel that enters 
in the Slavnov-Taylor identity satisfied by the full three-gluon vertex~\cite{Ball:1980ax}. 
In addition, $H_{\mu\nu}$  is related to the full gluon-ghost vertex, $\Gamma_\mu$, whose tensorial structure is given by
\be
-{\Gamma}_{\mu} =  B_1 p_{\mu} + B_2 k_{\mu},
\label{Gtens}
\ee
where $B_i=B_i(-k-p,k,p)$, 
with $k$ representing the momentum of the gluon and $p$ the one of the anti-ghost. Specifically, 
\be
p^\nu H_{\mu\nu}(-k-p,k,p)=-i\Gamma_{\mu}(-k-p,k,p).
\label{qH}
\ee
At tree-level, $H_{\mu\nu}^{(0)} = ig_{\mu\nu}$, and $\Gamma^{(0)}_\mu =-p_\mu$.

It turns out that the $1+G(p^2)$ and $L(p^2)$ defined in \1eq{LDec}
are related to $F(p^2)$ by an 
exact relation (valid in Landau gauge only)~\cite{Aguilar:2009pp}
\be
F^{-1}(p^2) = 1+G(p^2) + L(p^2).
\label{funrel}
\ee

\section{The universal and renormalization-group invariant part of the BS kernel} 

Let us consider the kernel appearing in 
a typical Bethe-Salpeter equation (BSE), shown in Fig.(\ref{fig:BSkernel}), 
which is contained in the gray box. To make contact with earlier works, we will divide it by a factor
of $4 \pi$, and will denote it by ${\cal K}$.
The kernel ${\cal K}$ receives a ``universal'' (process-independent) contribution, 
whose origin is the pure gauge sector of the theory; in that sense, this contribution 
constitutes the 
common ingredient of any such kernel, regardless of the nature of the particles 
between which it is embedded.

\begin{figure}[!t]
\includegraphics[scale=.7]{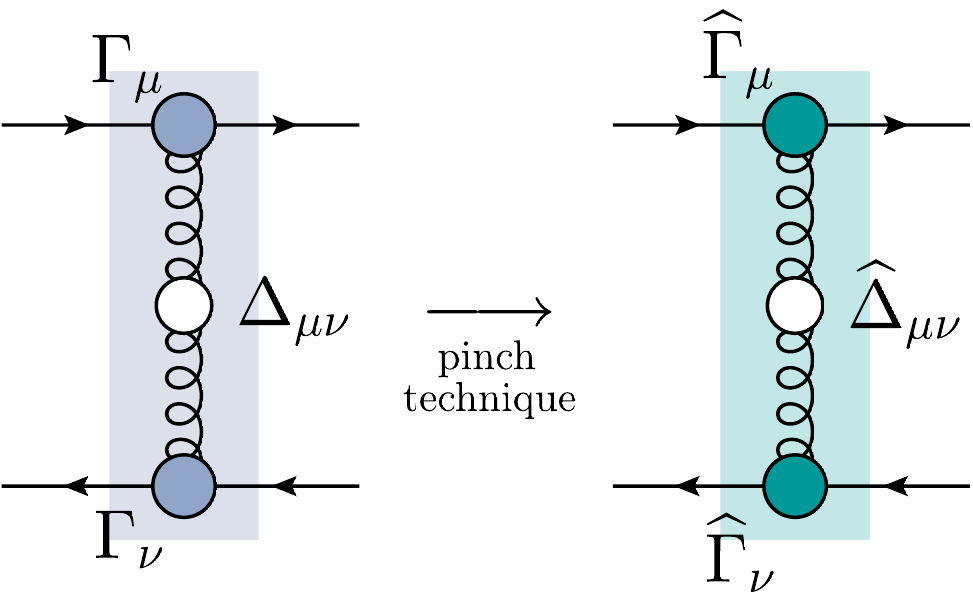}
\caption{\label{fig:PTkernel}The one-gluon exchange kernel before and after the pinch-technique rearrangement}
\end{figure}

The systematic diagrammatic identification of the precise pieces that constitute this particular 
quantity may be carried out following the procedure known in the literature as pinch technique (PT)~\cite{Binosi:2009qm}.
In general, the upshot of this construction is the rearrangement of a physical amplitude 
into sub-amplitudes with very special properties; in particular, one obtains  
vertices that satisfy QED-like Ward identities and a
gluon propagator that captures all the RG logarithms of the theory, see Fig.(\ref{fig:PTkernel}). In fact, it turns out 
that these latter quantities coincide precisely with the corresponding 
vertices and gluon propagator defined in the Background Field Method (BFM)~\cite{Abbott:1980hw}.
This particular identification persists both perturbatively, to all orders, as well as 
nonperturbative, at the level of the corresponding DSEs. 
 
In the case of the gluon propagator, the standard $\Delta(p^2)$ defined in \1eq{Delta}, and the 
scalar cofactor of the  PT-BFM gluon propagator, denoted by $\widehat\Delta (p^2)$,
are related by the exact relation~\cite{Binosi:2002ez} 
\be
\Delta(p^2) = \widehat\Delta(p^2) [1+G(p^2)]^2.
\label{BQI}
\ee
At the one-loop level, and keeping only UV logarithms, one has~\cite{Aguilar:2008xm}
\begin{align}
	1+G(p^2) &= 1 +\left(\frac{9}{4}\right)\frac{\alpha_s C_{A}}{12\pi} \ln\left(\frac{p^2}{\mu^2}\right);&
	\Delta^{-1}(p^2) &= p^2\left[1+ \left(\frac{13}{2}\right)\frac{\alpha_s C_{A} }{12\pi}\ln\left(\frac{p^2}{\mu^2}\right)\right], 
\label{pert_gluon}
\end{align}
and thus
\be
\widehat\Delta^{-1}(p^2) = p^2 \left[1+ b \alpha_s\ \ln\left(\frac{p^2}{\mu^2}\right)\right],
\ee
where $b = 11 C_A/12\pi$ is the first coefficient of the Yang-Mills  $\beta$ function, as it should~\cite{Abbott:1980hw}.

Similarly,
the PT-BFM quark-gluon vertex $\widehat\Gamma^a_\mu=\frac{\lambda^a}2\widehat\Gamma_\mu$, which satisfies the QED-like Ward identity
\be
q^\mu \widehat{\Gamma}_{\mu}(q,p_2,-p_1)= S^{-1}(p_1) - S^{-1}(p_2),  
\label{WI}
\ee
is related to the conventional $\Gamma_\mu$ by the BQI
\be
[1+G(q^2)]\Gamma_\mu(q,p_2,-p_1) =\widehat{\Gamma}_\mu(q,p_2,-p_1)+S^{-1}(p_1)\Q{\mu}(q,p_2,-p_1)+\oQ{\mu}(-q,p_1,-p_2)S^{-1}(p_2),
\label{BQI-final}
\ee
where $S^{-1}(p)$ is  the inverse of the full quark propagator, with $S^{-1}(p) = A(p^2)\,\sla{p} - B(p^2)$,
and the quantities $\Q{\mu}$ and $\oQ{\mu}$ are auxiliary three-point functions containing composite vertices.
The important point for what follows is that the last two terms on the rhs of \1eq{BQI-final} vanish when the external quarks are 
on shell; otherwise, they cancel against other (process-dependent) contributions.

To see how these considerations apply to the case at hand, let us express ${\cal K}$
in terms of the basic field-theoretic quantities that comprise it, namely (suppressing all spinor indices)
\be
{\cal K}(p,q_+,-q_-) =  \alpha_s \Gamma_\mu(p,q_{+},-p-q_+) \textstyle{\left(\frac{\lambda^a}2\right)}\Delta^{\mu\nu}(p)\textstyle{\left(\frac{\lambda^a}2\right)} \Gamma_\nu(-p,-q_{-},p+q_-),
\label{fullK}
\ee
where $\alpha_s = g^2/4\pi$, and $g$ is the gauge coupling.

As a consequence of \1eq{BQI-final} and \1eq{BQI},  \1eq{fullK} may be cast into the equivalent form 
\be
{\cal K}(p,q_+,-q_-) =  \alpha_s \widehat\Gamma_\mu(p,q_{+},-p-q_+) \textstyle{\left(\frac{\lambda^a}2\right)}\widehat\Delta^{\mu\nu}(p)\textstyle{\left(\frac{\lambda^a}2\right)} \widehat\Gamma_\nu(-p,-q_{-},p+q_-).
\label{fullK1}
\ee
In what follows we will focus only on the 
the part of $\widehat\Gamma_\mu$ that is  
proportional to $\gamma_\mu$, namely  \mbox{$\widehat\Gamma_\mu = \gamma_\mu \widehat\Gamma_1 +\cdots$}, so that the corresponding contribution to the ${\cal K}$ of \1eq{fullK1}, to be denoted by  
${\cal C}$, is given by 
\be
{\cal C}(p,q_+,-q_-) = \underbrace{\alpha_s \widehat\Delta(p^2)}_{\rm universal}  
\underbrace{[\widehat\Gamma_1(p,q_{+},-p-q_+) (\gamma_\mu)\textstyle{\left(\frac{\lambda^a}2\right)}  P_{\mu\nu}(p)\textstyle{\left(\frac{\lambda^a}2\right)} (\gamma_\nu)\widehat\Gamma_1(-p,-q_{-},p+q_-)]}_{\rm process-dependent},
\label{Kgg}
\ee

Due to the special Ward identities satisfied by the PT-BFM Green's functions [{\it e.g.}, \1eq{WI}], 
the (dimensionful) universal combination~\cite{Aguilar:2009nf}
\be
{\widehat d}(p^2) = \alpha_s \widehat\Delta(p^2)
= \frac{\alpha_s \Delta(p^2)}{\left[1+G(p^2)\right]^2},
\label{defd}
\ee
introduced in \1eq{Kgg}, is also renormalization-group invariant (RGI).
Indeed, since $g(\mu^2) =Z_g^{-1}(\mu^2) g_0$ and 
\mbox{$\widehat\Delta(p^2,\mu^2) =  \widehat{Z}^{-1}_A(\mu^2)\widehat{\Delta}_0(p^2)$}, 
where the ``0'' subscript indicates bare quantities, and the 
QED-like relation ${Z}_{g} = {\widehat Z}^{-1/2}_{A}$ is valid, we have that 
\be
{\widehat d}_0(p^2) = {\alpha_s}_0 \widehat\Delta_0(p^2) = \alpha_s(\mu^2) \widehat\Delta(p^2,\mu^2) = {\widehat d}(p^2)
\label{ptrgi}
\ee
maintains the same form before and after renormalization, {\it i.e.}, it 
forms a RGI ($\mu$-independent) quantity.

\section{Nonperturbative evaluation of ${\widehat d}(p^2)$}

In this section we use a combination of lattice results and dynamical equations to determine 
the nonperturbative form of the fundamental quantity ${\widehat d}(p^2)$ given in \1eq{defd}.
To that end, we will use for $\Delta(p^2)$ the lattice data from~\cite{Bogolubsky:2009dc}, whilst for 
the quantities $1+G(p^2)$ and $L(p^2)$ we will use the following set of (renormalized) equations 
\bea
1+G(p^2)\! &=&\! Z_c + \frac{g^2 C_{\rm {A}}}{d-1}
\int_k\! \left[(d-2) + \frac{(k \cdot p)^2}{k^2 p^2} \right] B_1(-k,0,k)\Delta (k)  D(k+p),
\nonumber\\
L(p^2)\! &=&\!  \frac{g^2 C_{\rm {A}}}{d-1}\!
\int_k\! \left[1- \frac{(k \cdot p)^2}{k^2 p^2}\right] B_1(-k,0,k) \Delta (k)  D(k+p),
%\nonumber\\
%F^{-1}(p^2) \! &=&\! Z_c  +g^2 C_{\rm {A}} \int_k\, \left[1- \frac{(k \cdot p)^2}{k^2 p^2}\right] B_1(-k,0,k)\Delta (k)  D(k+p),
\label{ttall}
\eea 
where the renormalization constant $Z_c$ is determined from the MOM condition $F(\mu^2)=1$. 
The assumptions and approximations employed in the derivation of the above set of equations have been 
explained in detail elsewhere~\cite{,Aguilar:2009pp,Aguilar:2009nf}. In addition, the 
form factor $B_1$ that enters in the three equations of (\ref{ttall})
has been computed from its own DSE, in the limit of vanishing ghost momentum~\cite{Aguilar:2013xqa}. 
Note that even though only  $G(p^2)$ enters into the definition of \1eq{defd}, from \1eq{ttall}
we will determine $G(p^2)$, $L(p^2)$, and $F(p^2)$; this in turn, will allow us to test numerically the validity of
\1eq{funrel}, which constitutes a nontrivial check of the entire numerical procedure.

\begin{figure}[!t]
\includegraphics[scale=0.95]{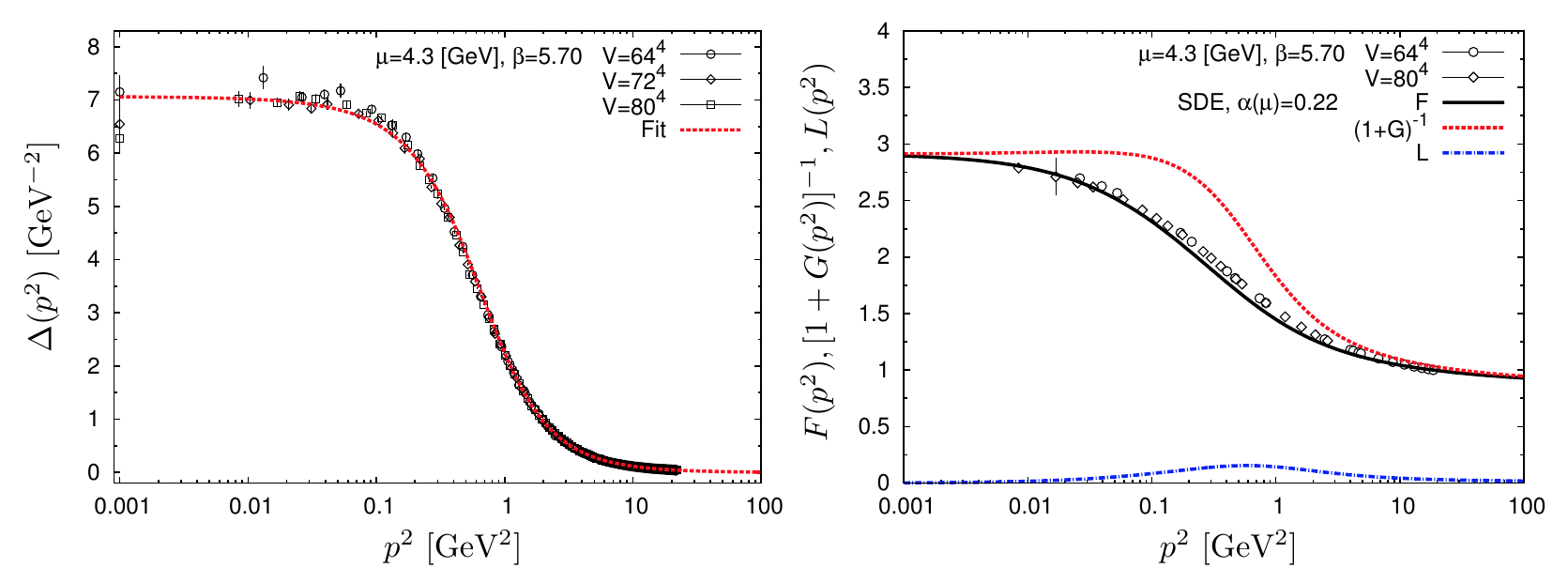}
\includegraphics[scale=0.95]{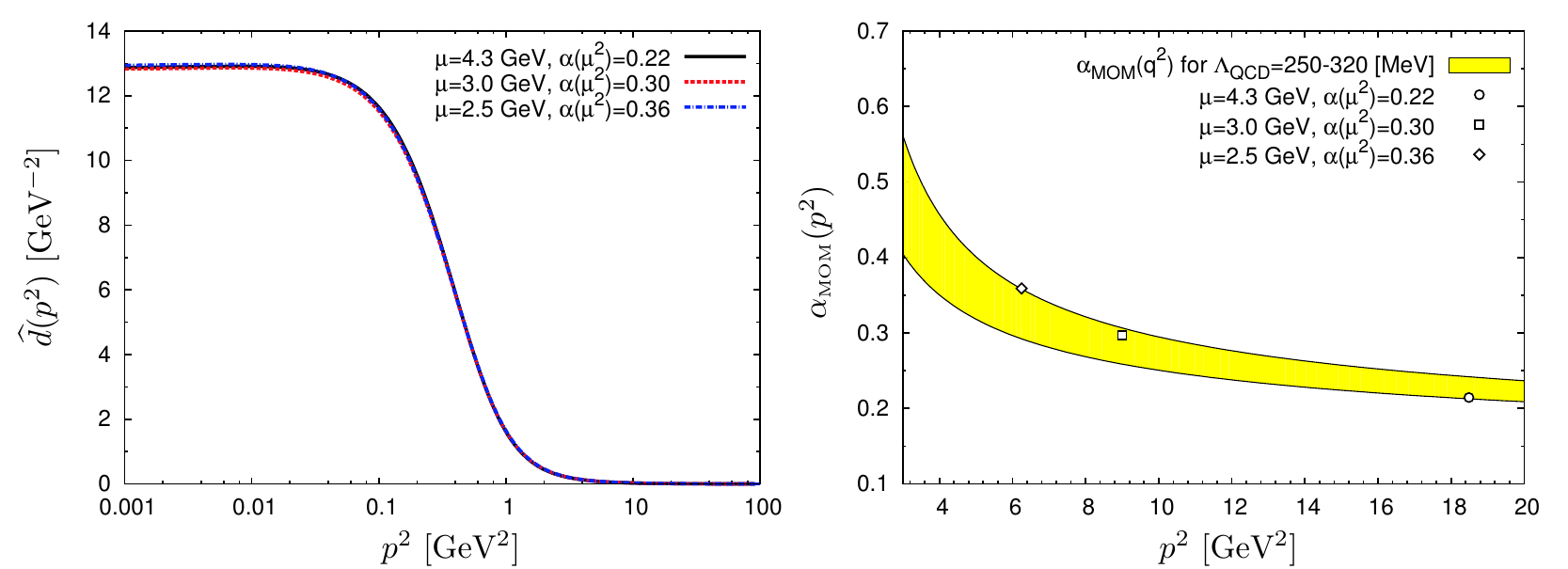}
\caption{\label{fig:input} The Landau gauge gluon propagator (top left panel) used for the numerical evaluation of the DSEs in ~\noeq{ttall} (top right panel) for the case $\mu=4.3$ GeV (lattice data are from~ of~\cite{Bogolubsky:2009dc}). The RGI quantity $\widehat{d}(p^2)$ evaluated at $\mu=4.3$, $3.0$ and $2.5$ GeV (bottom left panel). The resulting curves are then rescaled by $\alpha_s(\mu^2)=0.22$, $0.30$ and $0.36$, respectively, in  compliance with values obtained for $\alpha_\s{\mathrm{MOM}}$ when $\Lambda_{\s{\mathrm{QCD}}}$ varies between 250 and 320 MeV (yellow band on the bottom right panel)~\cite{Boucaud:2008gn}.}
\end{figure}

The results of the numerical solution of \1eq{ttall} are shown in the right panel of~\fig{fig:input}.
It is clear that the solutions obtained for $1+G$ and $L$ satisfy the fundamental relation~\noeq{ttall} at a high level of accuracy; this provides a non-trivial test for the integration routines used in solving the ghost DSE.
Note also that even though $L(p^2)$ vanishes at the origin, 
it has a non-vanishing support in the region of physical 
interest (see the top right panel of~\fig{fig:input}).

The quantity $\widehat d(p^2)$ is RGI, as is evident from the bottom left panel of~\fig{fig:input}. There, $\widehat d(p^2)$ is shown when evaluated at three different renormalization points: $\mu=4.3$, $3.0$ and $2.5$ GeV, for which the corresponding coupling reads $\alpha_s(\mu^2)=0.22$, $0.30$ and $0.36$ respectively. 
As can be seen in the bottom right panel, these values for the strong couplings are in very good agreement 
with those obtained from detailed calculations of the gauge 
coupling, $\alpha_\s{\mathrm{MOM}}$, 
in the momentum subtraction (MOM) scheme, for values of $\Lambda_\s{\rm QCD}$ between 250 and 320 MeV~\cite{Boucaud:2008gn}.

\section{Comparison between ``top-down'' and ``bottom-up'' approaches} 

%%%%%%%%%%%%%%%%%%%%%%%%%%%%%%%%%%%%%%%%%%%%%%%%%%%%%%%%%%%%%%%%%%%%%%%%%%%%%%%%%%%%%%%%%%%%%%
In order to make contact with the relevant literature on BSEs, we note that the quantity ${\cal C}$ appearing in~\1eq{Kgg} has been traditionally cast in the form
\begin{equation}
{\cal C}(p,q_+,-q_-) =\frac{{\cal I}(p^2)}{p^2}\otimes (\mathrm{process-dependent}), 
\label{Aeff}
\end{equation}
where ${\cal I}(p^2)$ is a dimensionless quantity, which may be interpreted as the effective interaction strength 
of the quark-gluon system.
Evidently, in the PT-BFM case 
one must extract an analogous quantity from the ${\widehat d}(p^2)$ of \1eq{defd} through multiplication by $p^2$, namely 
\be
{\cal I}(p^2) =  p^2 {\widehat d}(p^2).
\label{Asde}
\ee

%%%%%%%%%%%%%%%%%%%%%%%%%%%%%%%%%%%%%%%%%%%%%%%%%%%%%%%%%%%%%%%%%%%%%%%%%

%%%%%%%%%%%%%%%%%%%%%%%%%%%%%%%%%%%%%%%%%%%%%%%%%%%%%%%%%%%%%%%%%%%%%%%%%

At this point, the ${\cal I}(p^2)$ obtained from \1eq{Asde} can be compared with the corresponding quantities defined in the bottom-up framework.
As reviewed elsewhere~(see, {\it e.g.}, \cite{Cloet:2013jya}), successful explanations and predictions of numerous hadron observables can be obtained by choosing  
\begin{equation}
{\cal I}(p^2)=p^2 {\cal G}(p^2);\quad
{\cal G}(p^2)=\frac{8\pi^2}{\omega^4}D{\mathrm e}^{-p^2/\omega^2}+\frac{8\pi^2\gamma_m (1-{\mathrm e}^{-p^2/4m_t^2})}{k^2\ln[\tau+(1+p^2/\Lambda_\s{\rm QCD}^2)^2]},
\label{bu}
\end{equation}
where $\gamma_m=12/(33-2N_f)$ [typically, $N_f=4$], $\Lambda_\s{\rm QCD}=0.57$ GeV (in the MOM scheme); \mbox{$\tau={\rm e}^2-1$}, $m_t=0.5$ GeV. Note that  
$D$ and $\omega$ are not independent, but must be related by 
$D\omega=(\varsigma_\s{\mathrm{G}})^3=\mathrm{const}$ and $\omega\in[0.4,0.6]$ GeV; then, 
one can reproduce a large body of observable properties of ground-state vector- and isospin-nonzero pesudoscalar mesons, as well as various properties of the nucleon and $\Delta$ resonance~\cite{Qin:2011dd}. The parameter $\varsigma_\s{\mathrm{G}}$ is 
fixed by requiring that  
the correct value of the pion decay constant $f_\pi$ is reproduced; its precise value depends on the vertex Ansatz employed. 

In the case of the rainbow-ladder (RL) truncation~\cite{Maris:1999nt}, corresponding to $\Gamma^\nu \sim\gamma_\nu$, one has~$\varsigma_\s{\mathrm{RL}}=0.87$ GeV. 
Instead, the improved truncation scheme of~\cite{Qin:2011dd}, 
which incorporates dynamical chiral symmetry breaking (DB) effects by using   
a more sophisticated representation for $\Gamma^\nu$, gives $\varsigma_\s{\rm DB}=0.55$ GeV.

\begin{figure}[!t]
\includegraphics[scale=0.58]{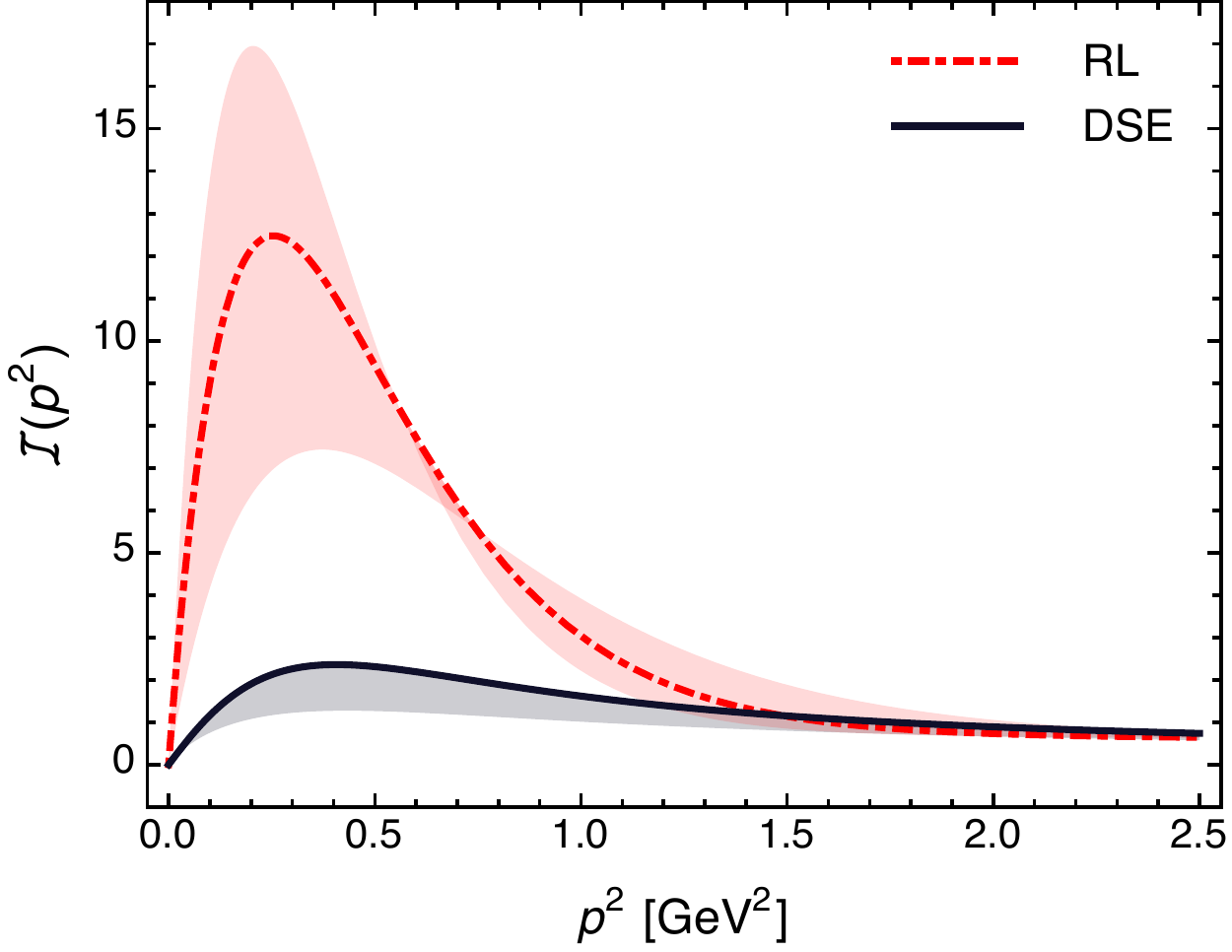}
\includegraphics[scale=0.574]{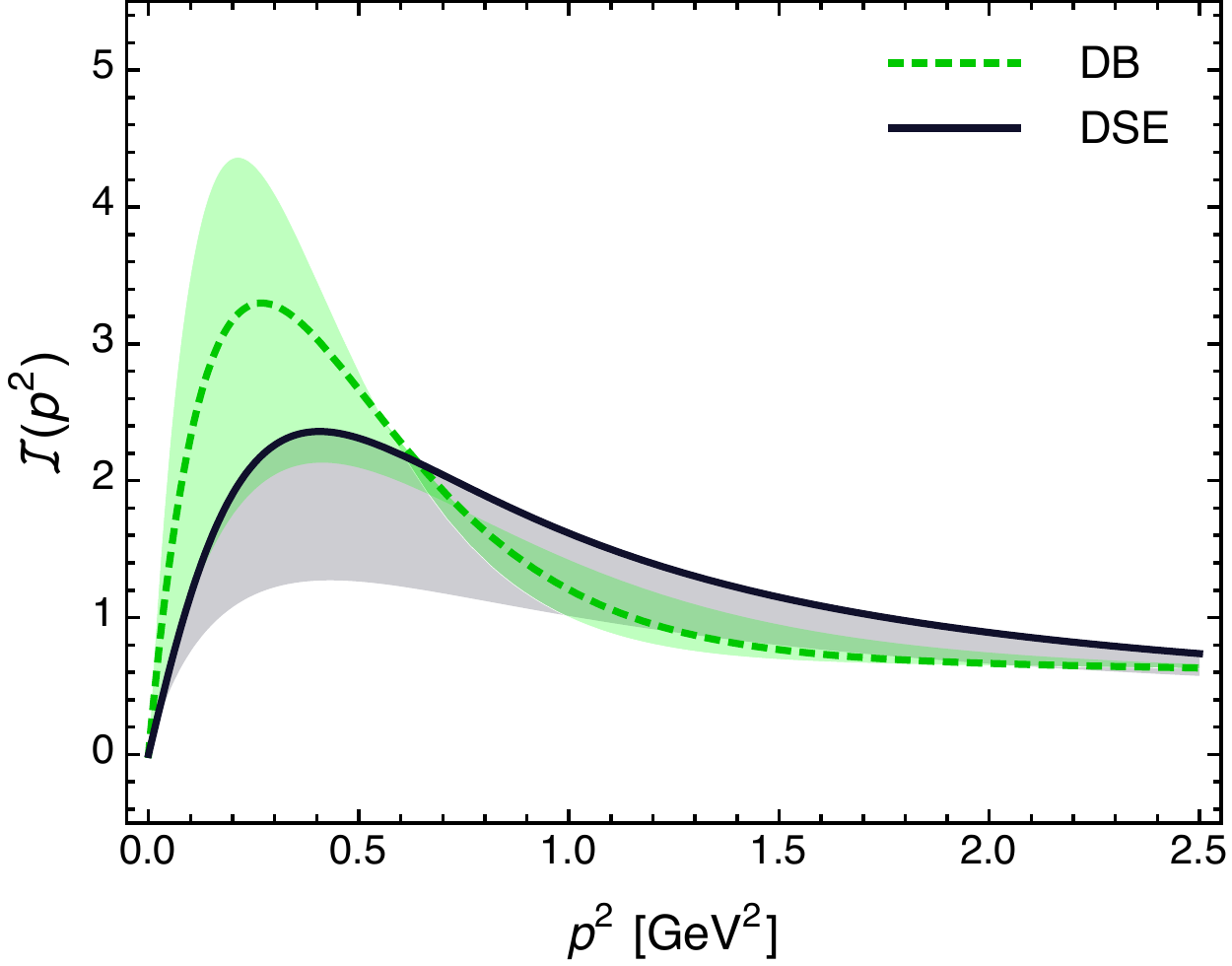}
\caption{\label{fig:direct-comp}Comparison of the interaction strength ${\cal I}$ evaluated in the RL, IRL and DSE schemes.
The shaded ares represent phenomenologically acceptable ranges.}
\end{figure} 

The results of the top-down approach of~\cite{Binosi:2014aea} 
(denoted by DSE), and the two bottom-up approaches (RL and DB) summarized above, are   
shown in Fig.~(\ref{fig:direct-comp}). It is clear that   
while the RL interaction strength is far larger than that of the DSE, the comparison between the 
DB and DSE results is very favorable. Evidently, the DB approach captures corrections that are not included 
in the RL, and is therefore much closer to the DSE result (see also~\cite{Binosi:2016rxz}).

\section{Conclusions}
In this work we have reviewed recent developments towards  
a first-principle derivation of the BSE 
kernel from the dynamical equations obtained from the QCD Lagrangian. This particular effort, in turn, 
 bridges to a large extent the gap between nonperturbative continuum QCD studies
and bound-state phenomenology.

\begin{acknowledgements}
This research is supported by the Spanish MEYC under 
grants FPA2011-23596, FPA2014-53631-C2-1-P, and SEV-2014-0398,
and the Generalitat Valenciana under grant “PrometeoII/2014/066”.
I would like to thank the organizers for their kind invitation and their 
warm hospitality during the workshop. 
\end{acknowledgements}

\end{document}